\documentclass[12pt, a4paper]{article}
 \usepackage[font=small,format=plain,labelfont=bf,up,textfont=normal,up,justification=justified,singlelinecheck=false]{caption}
\usepackage[a4paper]{geometry}
\usepackage{hyperref}
\usepackage{amsfonts} 
\usepackage{amsmath} 
\usepackage{setspace}   
\usepackage{color} 
 
\usepackage[utf8,applemac]{inputenc} 
\usepackage{tensor}
\usepackage{cite} 
\usepackage{tikz}
\usepackage{graphicx}

\usepackage{dcolumn}
\usepackage{bm}

\newcommand{\bea}{\begin{eqnarray}}
\newcommand{\eea}{\end{eqnarray}}
\newcommand{\be}{\begin{equation}}
\newcommand{\ee}{\end{equation}}
\def\nn{\nonumber}

  \newcommand{\beqs}{\begin{eqnarray}}
\newcommand{\eeqs}{\end{eqnarray}}

\title{Einstein black holes have finite classical hair}

\begin{document}


\setcounter{tocdepth}{1}

\begin{titlepage}

\begin{flushright}\vspace{-3cm}
{\small
\today }\end{flushright}
\vspace{0.2cm}

\begin{center}



{{ \LARGE{\bf{Bulk supertranslation memories:\vspace{0mm}\\ a concept reshaping the vacua and \vspace{2mm}\\ black holes of general relativity
}}}} 


\vspace{18mm}
\centerline{\large{\bf{Geoffrey Comp\`{e}re}}}

\vspace{2mm}
\normalsize
\bigskip\medskip
\textit{Universit\'{e} Libre de Bruxelles and International Solvay Institutes\\
CP 231, B-1050 Brussels, Belgium\\
e-mail: gcompere@ulb.ac.be
}

\vspace{20mm}

\begin{abstract}
\noindent 
{The memory effect is a prediction of general relativity on the same footing as the existence of gravitational waves. 
The memory effect is understood at future null infinity as a transition induced by null radiation from a Poincar\'e vacuum to another vacuum. Those are related by a supertranslation, which is a fundamental symmetry of asymptotically flat spacetimes.
In this essay, I argue that finite supertranslation diffeomorphisms should be extended into the bulk spacetime consistently with canonical charge conservation. It then leads to fascinating  geometrical features of gravitational Poincar\'e vacua. I then argue that in the process of black hole merger or gravitational collapse, dramatic but computable memory effects occur. They lead to a final stationary metric which qualitatively deviates from the Schwarzschild metric. 
}

\end{abstract}
\vspace{20mm}
{{ \it{Essay written for the Gravity Research Foundation \\ 2016 Awards for Essays on Gravitation. Honorable mention. }}} \vspace{10mm}


\end{center}

\end{titlepage}

\newpage

What is gravity? In general relativity, the gravitational field is the metric field. It contains the Newtonian potential corrected by relativistic effets, which leads to gravitational attraction, the bending of light and many other well-known gravitational phenomena. The metric also describes gravitational waves, a spin 2 field best isolated in linearized Einstein gravity, which couples to the Newtonian field at the non-linear level. Finally, the metric also contains a third physically distinct component: the \emph{supertranslation memory field}, originally defined close to future null infinity. This field is intimately coupled to gravitational waves but is related to a physically distinct effect: the memory effect \cite{Zeldovich:1974aa,Christodoulou:1991cr}. I will argue that the supertranslation memory field leads to new phenomena when considered in the bulk spacetime. 

{\vspace{12pt}\noindent \bf Gravitational memory \vspace{12pt}}

After the passage of either gravitational waves or null matter between two detectors placed in the asymptotic null region, the detectors generically acquire a finite relative spatial displacement and a finite time shift. This is the \emph{memory effect}, respectively non-linear for gravitational waves and linear for null matter \cite{Zeldovich:1974aa,Christodoulou:1991cr}. Since this displacement is a zero frequency effect, it cannot straightforwardly be detected by ground-based interferometers such as LIGO mainly because of seismic noise which blurs the signal at low frequencies $< 30 Hz$. Nevertheless, sophisticated data analyses might make such a measurement possible in the future \cite{Lasky:2016knh}. 

Memory effects do not exist in Newtonian gravity. In Einstein gravity, they naturally arise in post-Newtonian wave forms created by an inspiraling binary system. Such wave forms depend upon the entire past history of the system \cite{Blanchet:1987wq}. These ``hereditary effects'' can be identified as the tail effect and the non-linear memory effect \cite{Blanchet:1992br}. The tail effect first appears at 1.5PN order and is attributed to backscattering of past gravitational waves. It will not concern us here. The non-linear memory effect first appears at 2.5PN order and gives rise to a net cumulative change in the wave form at \emph{leading} Newtonian order after integrating over the past history of the source which effectively decreases the order by 2.5PN \cite{Arun:2004ff}. 
The fundamental nature of this leading order effect is rooted in the symmetry structure of Einstein gravity which I will describe next.

{\vspace{12pt}\noindent \bf Supertranslation symmetry and degenerate Poincar\'e vacua  \vspace{12pt}}

Another feature of Einstein gravity without cosmological constant, found earlier by Bondi, Metzner, van der Burg and Sachs \cite{Bondi:1962px,Sachs:1962wk}, is the existence of an infinite-dimensional asymptotic symmetry group at future and past null infinity: the so-called BMS group. The BMS group includes the Poincar\'e group but it extends the translation normal subgroup into a larger abelian normal subgroup.  The additional group elements are called the \emph{supertranslations}.  There is only one BMS symmetry of asymptotically flat gravity which acts on both future and past null infinity \cite{Strominger:2013jfa}. 

This symmetry has multiple consequences. The metric is not only determined around future null infinity by its asymptotic Newtonian potentials, but also by the value of the field at future null infinity which linearly shifts under supertranslation diffeomorphisms. This field is the \emph{supertranslation field}, usually denoted as $C(\theta,\phi)$, with the same arbitrary angle-dependence as a generic supertranslation. It is the Goldstone boson of spontaneously broken supertranslation invariance\cite{He:2014laa}. Since the supertranslations are abelian and commute with time translations, shifting the supertranslation field does not affect the energy. As a consequence, \emph{the Poincar\'e vacuum is degenerate in Einstein gravity} \cite{He:2014laa}. 

{\vspace{12pt}\noindent \bf Bulk supertranslation field from superrotation charge conservation \vspace{12pt}}

We define the Poincar\'e vacua as the metrics obtained by acting on Minkowski spacetime with the exponentiation of an infinitesimal supertranslation diffeomorphism. The definition of the diffeomorphism in the bulk spacetime is uniquely determined after fixing a convenient gauge. I now argue that a fundamental requirement on the gauge condition is that in stationary spacetimes \emph{all conserved canonical charges defined at infinity should be conserved upon shrinking the boundary sphere towards the bulk}. This requirement formally amounts to the existence of a phase space with symplectic symmetries, which extends the concept of asymptotic symmetries into the bulk \cite{Compere:2014cna,Compere:2015bca,Compere:2015knw} (see also \cite{Barnich:2010eb}). The supertranslation field is then promoted to a bulk concept, characterized by its conserved charges. 

Both static radial gauge and BMS gauge lead to such a phase space for the vacua \cite{Compere:2016jwb}. The metric can be built explicitly and reads in static radial gauge as
\begin{align}\label{vacua} \begin{split}
ds^2 &= - d(t+C_{(0,0)})^2 +  d\rho^2 + g_{AB} d\theta^Ad\theta^B,  \\
g_{AB} &= (\rho-C+C_{(0,0)})^2 \gamma_{AB} -2(\rho - C) D_A D_B C + D_A D_E C D_B D^E C
\end{split}\end{align}
where $\theta^A = (\theta,\phi)$, $\gamma_{AB}$ is the unit sphere metric and $D_A$ its covariant derivative. 
The lowest spherical harmonic of $C$ is denoted as $C_{(0,0)}$. This remarkably simple metric reveals a new feature of supertranslations: they are all spatial, except the zero mode which corresponds to time shifts. This feature cannot be observed at future null infinity which mixes time and space. 

It can be proved that all Poincar\'e charges are identically zero \cite{Compere:2016jwb}. However, other canonical surface charges can be defined: the superrotations, which extend the Lorentz charges \cite{Barnich:2010eb,Barnich:2011mi}. (Before more sophisticated phase space methods were developped \cite{Lee:1990nz,Iyer:1994ys,Barnich:2001jy,Barnich:2007bf}, precursors of such superrotations were called linkages \cite{Geroch:1981ut}). Such superrotations are finite and non-vanishing \cite{Flanagan:2015pxa,Compere:2016jwb} and therefore allow to distinguish the vacua. Their presence demonstrates the existence of a bulk defect: a large sphere cannot be shrunk to zero size otherwise such charges would be zero. 

A feature of these vacua is that the static radial gauge breaks down at  the (gauge dependent notion of) \emph{supertranslation horizon} $\rho = \rho_{SH}(C)$ where $\text{Det}(g_{AB}) = 0$ beyond which other coordinates are necessary. The nature of the source of superrotation charges, which lays beyond the supertranslation horizon, is unclear to me at this time. The shape of the supertranslation horizon is depicted in simple examples in Figure 1.

\begin{figure}[!htb]
\begin{minipage}{0.47\textwidth}
\centering
\includegraphics[scale=0.40]{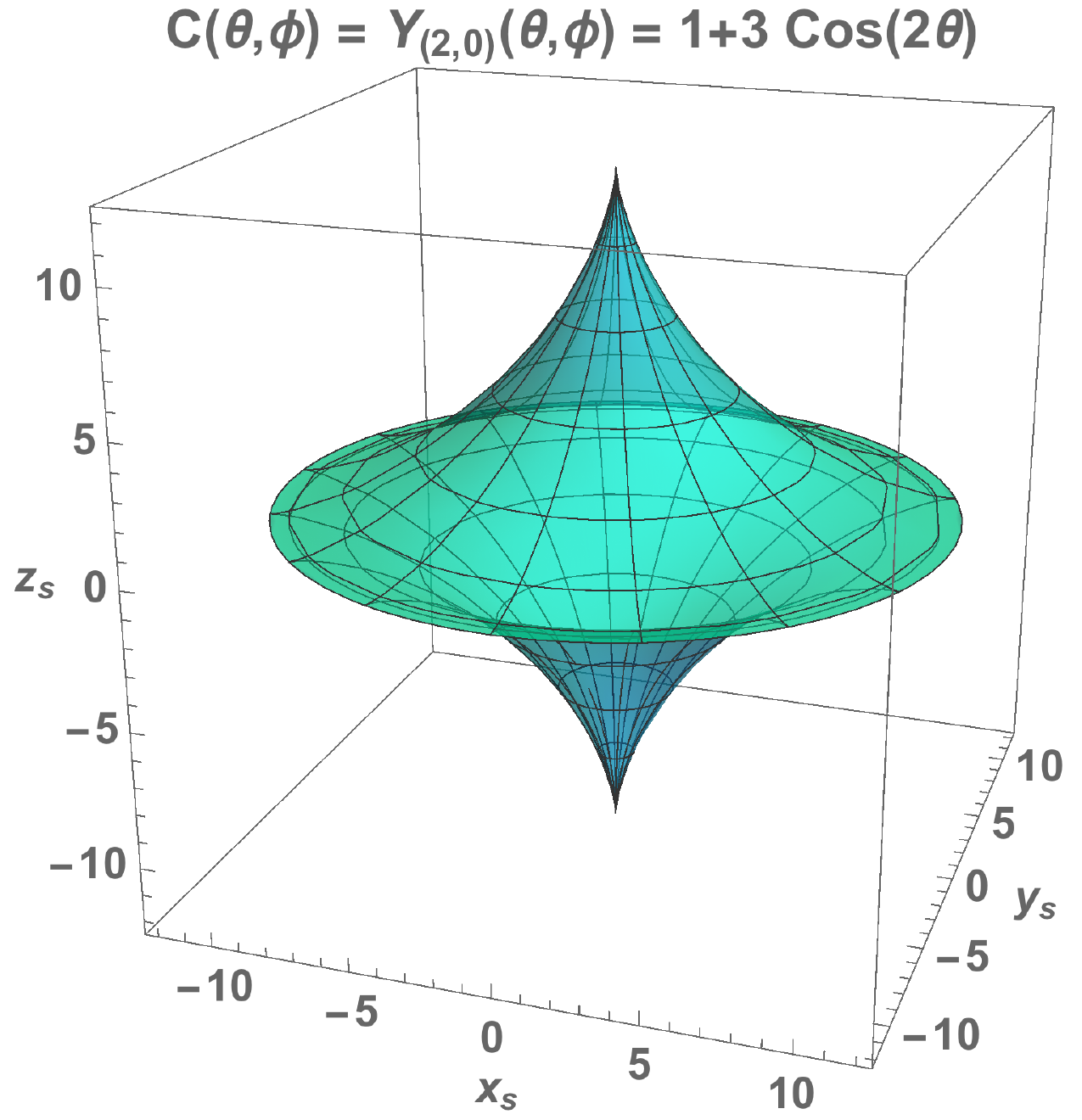}
\end{minipage}\hfill
\begin{minipage}{0.47\textwidth}
\centering
\includegraphics[scale=0.47]{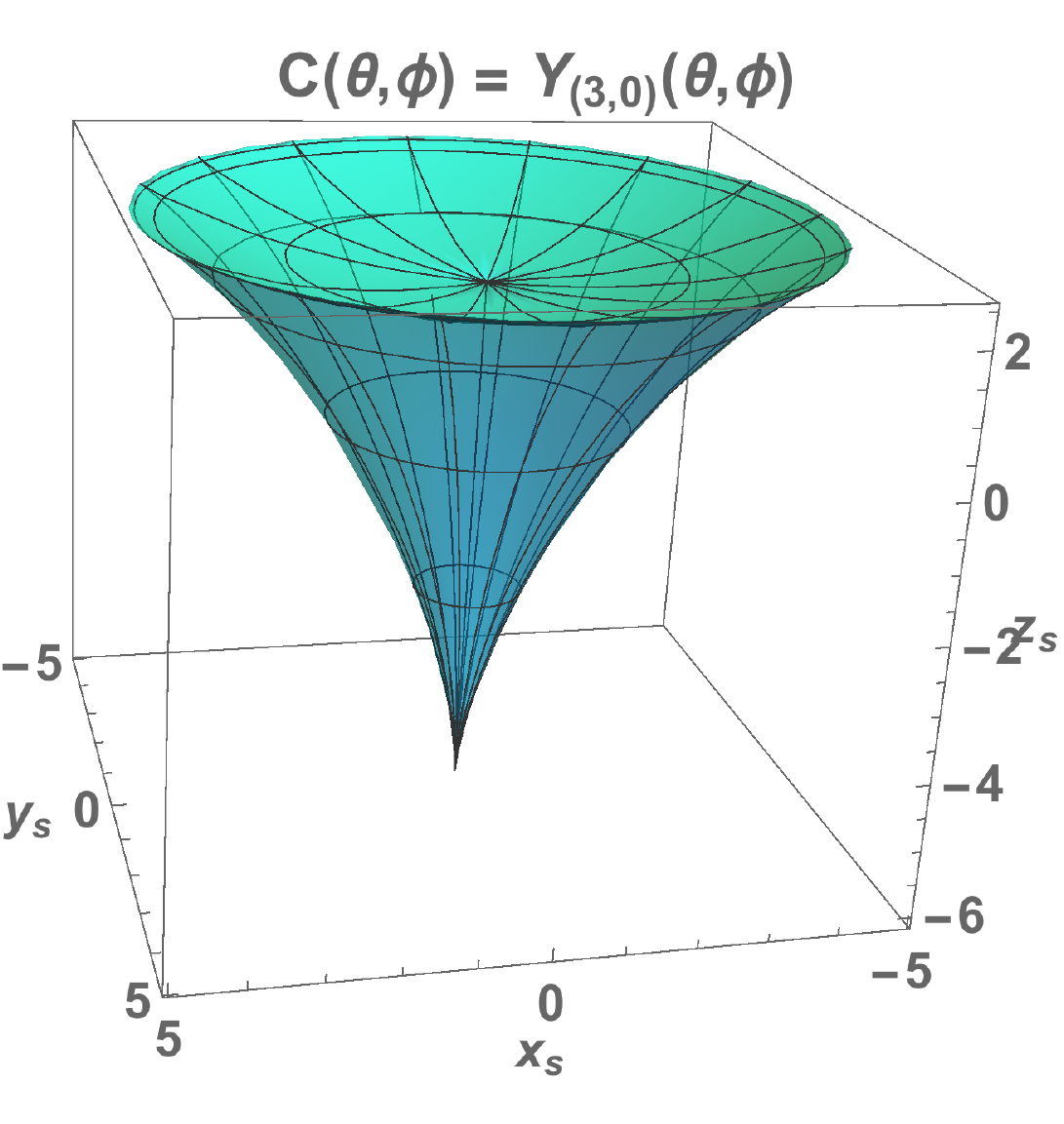}
\end{minipage}
\caption{Isometric embedding of the supertranslation horizon in Euclidean space $(x_s,y_s,z_s)$. The supertranslation field $C$ is either the $l=2$ (left) or $l=3$ (right) $m=0$ spherical harmonic.}
\end{figure}

{\vspace{0pt}\noindent \bf The global angle-dependent energy conservation law \vspace{12pt}}

The relationship between memory effects close to future null infinity and the supertranslation field was understood some time ago from an angle-dependent energy conservation law \cite{Geroch:1981ut,0264-9381-9-6-018} (see also \cite{Strominger:2014pwa}).  Einstein's equations imply that the difference of supertranslation field between the distinct vacua at retarded times $u_2$ and $u_1$ is related to the passage of radiation between $u_1$ and $u_2$ as
\bea
-\frac{1}{4}D_A D^A (D_A D^A+2) (C|_{u_2} - C|_{u_1} ) = m|_{u_2} - m|_{u_1}  +  \int_{u_2}^{u_1} du T_{uu}. \nn
\eea
Here $m$ is the Bondi mass and $T_{uu}$ is the sum of $1/4$ the Bondi news squared sourcing non-linear memory and $4\pi G$ times the matter null energy, sourcing linear memory. 
The change of the supertranslation field due to radiation precisely accounts for the relative displacement of inertial detectors. Therefore, memory effects can be understood as a change of Poincar\'e vacuum caused by the passage of radiation. 

Now, memories accumulate as well at past null infinity where a similar equation holds. A junction condition at spatial infinity is necessary to formulate a global conservation law between an initial state and a final state which takes both incoming and outgoing radiation into account. The junction condition was recently understood for small non-linear perturbations of Minkowski spacetime to amount to an antipodal map of future and past fields at spatial infinity \cite{Strominger:2013jfa}. After taking into account boundary conditions at early and late times \cite{Christodoulou:1993uv}, the \emph{global angle-dependent energy conservation law} between stationary initial and final states then reads as \cite{Strominger:2014pwa}
\bea
&&-\frac{1}{4} D_A D^A (D_A D^A+2)  (C|_{final}(\theta,\phi) - \Delta C|_{i_0} - C|_{initial}(\pi-\theta,\phi+\pi) )\nn \\
&&  = m|_{final} - m|_{in}  +  \int_{-\infty}^{final} du T_{uu}(\theta,\phi) -   \int_{initial}^{+\infty} dv T_{vv}(\pi-\theta,\phi+\pi) \label{eq4}. 
\eea
where $u$ ($v$) is retarded (advanced) time at future (past) null infinity. Here $ \Delta C|_{i_0}$ is the difference between $C(\theta,\phi)$ at the past of future null infinity and $C(\pi-\theta,\phi+\pi)$ at the future of past infinity. The lesson of this equation is that entire past history of the energy fluxes determines what the supertranslation field $C(\theta,\phi)$ is in the final state, up to a boundary term at spatial infinity which is not totally understood. 

{\vspace{12pt}\noindent \bf Black hole memories \vspace{12pt}}

What is the supertranslation field of astrophysical black holes, such as the final state of binary black hole merger GW150914 \cite{Abbott:2016blz} or Sagittarius A*? How does it compare with the mass $M$? Since all physical processes involved in merging, accretion and collapse are highly non-spherically symmetric, the $\ell \geq 2$ spherical harmonics in the right-hand side of \eqref{eq4} will be of order of the mass. Assuming $C|_{initial} =  \Delta C|_{i_0} = 0$ (which would have to be assessed) and after solving for the differential operator, the supertranslation field will be $O(M)$. 

Assuming zero angular momentum, the final black hole is constrained by uniqueness theorems to be diffeomorphic to the Schwarzschild black hole \cite{Hawking:1973uf,Carter:1971zc,Robinson:1975bv,Chrusciel:2008js,Alexakis:2009ch}. However, the diffeomorphism might be singular inside the Killing horizon and act as a source for the supertranslation field. The black hole final state will then qualitatively differ from the Schwarzschild black hole because of the presence of additional conserved charges: the superrotations. Exploiting this loophole, the static metric was recently constructed \cite{Compere:2016hzt}. It simply reads in static radial gauge as
\bea
ds^2 &=& - \frac{\left( 1-\frac{M}{2\rho_s}\right)^2}{\left( 1+\frac{M}{2\rho_s}\right)^2}dt^2 + \left( 1+\frac{M}{2\rho_s}\right)^4 \Big( d\rho^2 + g_{AB} d\theta^Ad\theta^B \Big) 
\label{newm}
\eea
where $\rho_s^2 = (\rho - C+C_{(0,0)})^2 + D_A C D^A C$ and $g_{AB}$ was given in \eqref{vacua}. It admits the same mass as the Schwarzschild black hole and can therefore be coined the Schwarzschild black hole with BMS soft hair (see also \cite{Hawking:2016msc}). The hair can be observed in principle by measuring the metric around the black hole and reconstructing the superrotation charges. The metric admits a coordinate supertranslation horizon as well as the usual Killing horizon which have interesting competiting effects, constrained by the null energy condition through \eqref{eq4} \cite{Compere:2016hzt}. 

{\vspace{12pt}\noindent \bf Conclusion \vspace{12pt}}

If one takes the BMS group as a fundamental symmetry of asymptotically flat Einstein theory \emph{in the bulk spacetime}, one is led to the existence of Poincar\'e vacua with new intringuing properties such as the supertranslation horizon and localized superrotation charges. The accumulation of memories leads to a final black hole metric which qualitatively differs from the Schwarzschild metric by a measurable, classical and $O(M)$ deviation, after taking into account the global angle-dependent  conservation of energy. These are some of the fascinating outcomes of bulk supertranslation memories.


\section*{Acknowledgments}

I acknowledge Jiang Long for our enjoyable and productive recent collaboration, and L. Donnay, K. Hajian, P.-H. Lambert, P.-J. Mao, W. Schulgin, A. Seraj and S. Sheikh-Jabbari  for our earlier collaboration on symplectic symmetries. I thank G. Barnich and J. de Boer for interesting discussions. I acknowledge the current support of the ERC Starting Grant 335146 HoloBHC ``Holography for realistic black holes and cosmology''. I am Research Associate of the Fonds de la Recherche Scientifique F.R.S.-FNRS (Belgium). This work is also partially supported by FNRS-Belgium (convention IISN 4.4503.15).
\appendix

\providecommand{\href}[2]{#2}\begingroup\raggedright\endgroup

\end{document}